\documentclass{nature}
\spacing{1}

\usepackage{graphicx}
\usepackage{amssymb}
\usepackage{amsmath}
\usepackage{gensymb}
\usepackage[square,sort,comma,numbers]{natbib}
\usepackage[switch]{lineno} 
\usepackage[hidelinks]{hyperref}
\usepackage{caption}
\usepackage{subcaption}

\newcommand{\la}{\,\rlap{\raise 0.4ex\hbox{$<$}}{\lower 0.8ex\hbox{$\sim$}}\,}
\newcommand{\ga}{\,\rlap{\raise 0.4ex\hbox{$>$}}{\lower 0.8ex\hbox{$\sim$}}\,}

\newcommand{\pcc}{\mathrm{pc}\,\mathrm{cm}^{-3}}
\newcommand{\pcm}{{pc\,cm$^{-3}$}}

\newcommand{\rmm}{{rad\,m$^{-2}$}}
\newcommand{\mr}{\mathrm}
\newcommand{\sr}{R_\odot}
\newcommand{\vp}{\vec{\mr{P}}}
\newcommand{\vom}{\vec{\Omega}}
\newcommand{\rhq}{\rho_Q}
\newcommand{\rhu}{\rho_U}
\newcommand{\rhv}{\rho_V}
\newcommand{\rhl}{\rho_L}
\newcommand{\bz}{\hat{B}_z}

\newcommand{\tz}{\tilde{z}}

\newcommand{\vs}{\vec{\mr{S}}}

\newcommand{\AFF}[1]{$^{\foreach\d[count=\ni]in{#1}{\ifnum\ni=1\ref{\d}\else,\ref{\d}\fi}}$}
\let\mr=\mathrm
\usepackage[colorinlistoftodos]{todonotes}

\title{A Highly Variable Magnetized Environment in a Pulsar Binary resembling Fast Radio Bursts}
\author{Dongzi~Li\AFF{aff:caltech}\thanks{E-mail:dongzili@caltech.edu,\href{https://orcid.org/0000-0001-7931-0607}{ \hspace{2mm} orcid.org/0000-0001-7931-0607 }},
Anna Bilous\AFF{aff:astron},
Scott Ransom\AFF{aff:nrao},
Robert Main\AFF{aff:mpifr}, 
Yuan-Pei Yang\AFF{aff:swifa}
}

\begin{document}

\maketitle
\begin{affiliations}
\item Cahill Center for Astronomy and Astrophysics, California Institute of Technology, 1216 E California Boulevard, Pasadena, CA 91125, USA \label{aff:caltech}
\item ASTRON, the Netherlands Institute for Radio Astronomy, Postbus 2, 7990 AA Dwingeloo, The Netherlands \label{aff:astron}
\item National Radio Astronomy Observatory, Charlottesville, VA, USA \label{aff:nrao}
\item Max-Planck-Institut f{\"u}r Radioastronomie, Auf dem H{\"u}gel 69, D-53121 Bonn, Germany \label{aff:mpifr}
\item South-Western Institute For Astronomy Research, Yunnan University, Yunnan 650504, P.~R.~China\label{aff:swifa}
\end{affiliations}
\begin{abstract}
Fast radio bursts (FRBs) are short, intense extragalactic radio bursts of unknown origin \cite{2007Lorimer}. Recent polarimetric studies have shown that a noticeable fraction of the repeating FRBs display irregular, short-time variations of the Faraday rotation measure (RM) \cite{Xu2021,Luo2020,2022Anna-Thomas}. Moreover, evidence for rare propagation effects such as Faraday conversion and polarized attenuation is seen in at least one FRB repeater \cite{Xu2021}. Together, they suggest a highly variable magneto-active circum-burst environment. 
In this paper, we report similar behavior in a globular cluster pulsar binary system PSR B1744$-$24A \cite{Lyne1990}. 
We observe irregular fast changes of RM with both signs at random orbital phases as well as profile changes of the circular polarization when the pulsar emission passes close to the companion. The latter provides strong evidence for Faraday conversion and circularly polarized attenuation.
These similarities between PSR B1744$-$24A and some FRB repeaters, as well as the possible binary-produced long-term periodicity of two active repeaters \cite{2020Chime,2020Rajwade}, and the discovery of a nearby FRB in a globular cluster\cite{2021Bhardwaj,2022Kirsten}, where pulsar binaries are common, all suggest that some fraction of FRBs have binary companions.

\end{abstract}

Located in the globular cluster Terzan 5, PSR B1744$-$24A (Ter5A) is a 11.56-ms ``redback'' pulsar in a 1.82-hr tight orbit with a $\sim$0.09$M_\odot$ companion \cite{Lyne1990}. For more than 15 years, it has been routinely observed with the 100-m Green Bank Telescope (GBT)  
with a cadence of a few months. 
The emission from Ter5A is usually eclipsed around pulsar superior conjunction, i.e. orbital phase $\Phi=0.25$, when the pulsar is behind the companion. Small irregular eclipses at other orbital phases and long eclipses spanning the full orbit are also common \cite{Lyne1990}. These features can usually last several orbits. 
While the detailed analysis of Ter5A's radio properties from the whole observational sample will be presented elsewhere, here we report on a couple of observations which show two interesting kinds of polarized propagation phenomena -  rapid RM variation at random orbital phases and a strong change in the circular polarization near superior conjunction. 

\begin{figure*}
  \centering
  \includegraphics[width=0.9\textwidth]{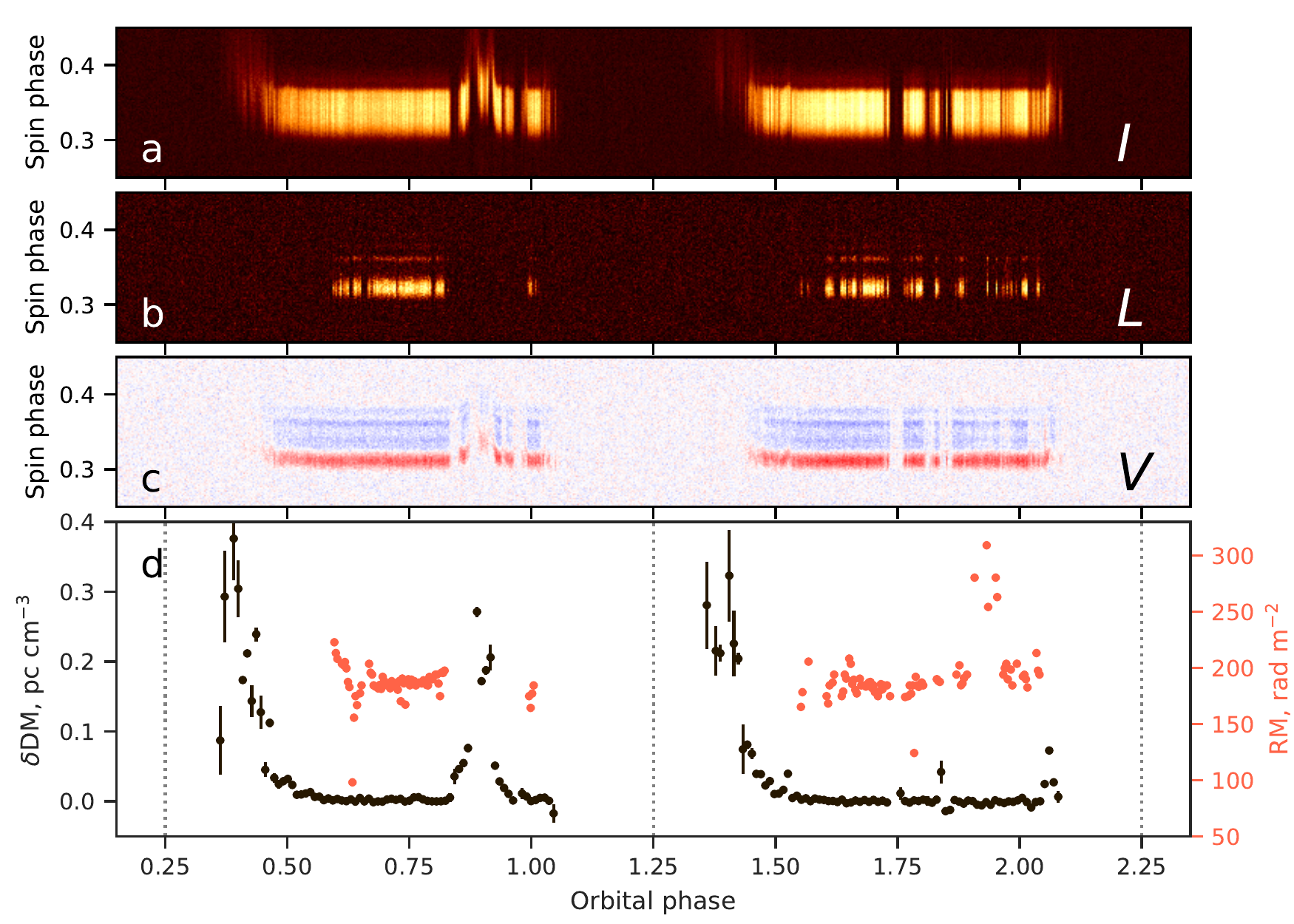}
  \includegraphics[width=0.9\textwidth]{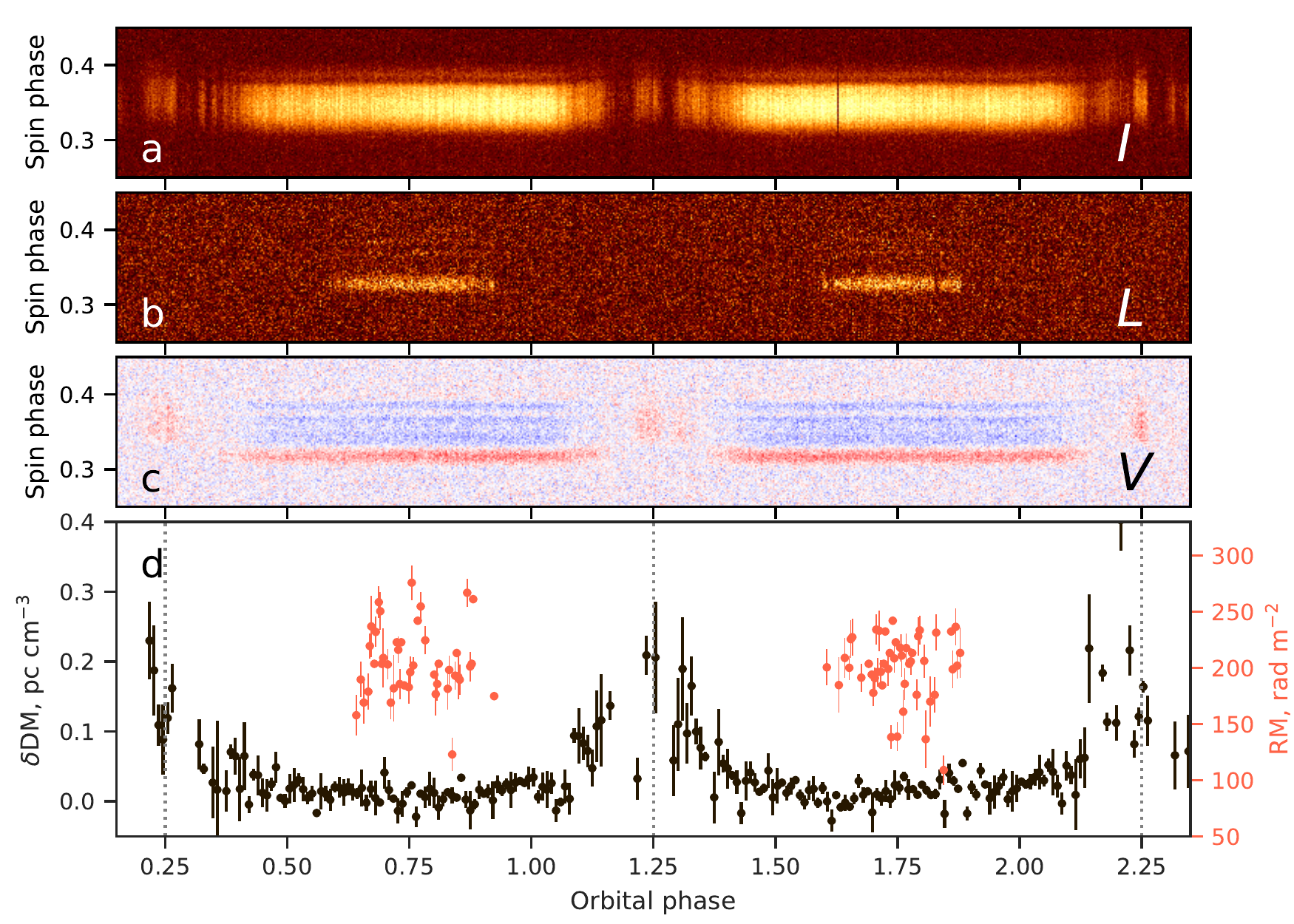}
      \caption{\textbf{Highly variable Ter5A polarization versus orbital phase.} \textbf{(a)} Pulse intensity ($I$) against orbital and spin phase at 1.5~GHz (top panel) and 2~GHz (bottom panel).  \textbf{(b)} Linear polarization ($L$) with Faraday rotation corrected using local RM value.
      \textbf{(c)} Circular polarization ($V$), with red/blue corresponding to positive/negative value. 
      Panel a,b,c are both dedispersed with constant $\mathrm{DM_0}=242.36~\pcc$, folded with 512 phase bins and averaged over 20-s.
      \textbf{(d)}
      $\delta \mathrm{DM} = \mathrm{DM}-\mr{DM}_0$ in 1-min integrations (black dots) and $\mathrm{RM}$ in 20-s integrations (red dots). The superior conjunction of the pulsar is indicated with dashed lines. Irregular, fast variations of RM of both signs, as well as depolarization are seen at random orbital phases due to the magnetized plasma from the companion.}
  \label{fig:obs_overview}
\end{figure*}

In Fig.~\ref{fig:obs_overview} left panel, the observation at 1.5\,GHz has shown a sudden 50\% decrease of RM at orbital phases of 0.65 and 1.8, which is $140\degree$ and $160\degree$ from superior conjunction, when the companion is almost behind the pulsar. 
At orbital phase 1.9, $\sim 120\degree$ away from superior conjunction, the RM has a sudden 50\% increase. Similar levels of RM variation are seen at 2~GHz, although with larger error bars (Fig~\ref{fig:obs_overview} right panel).
No significant dispersion measure (DM) variation up to $0.01$\,\pcm\ is detected during the RM jumps, 
implying a lower limit of the average parallel magnetic field in the material of the companion wind $\langle B\rangle=
12~\mr{mG}\, (\Delta\mr{RM}/100~\mr{rad}~\mr{cm}^{-2})/(\Delta\mr{DM}/0.01~\pcc)>12~\mr{mG}$.
Large RM variation of $>3000$\,{\rmm} is expected at $\Phi=0.9$ from this lower limit of magnetic field, because significant $\Delta$\,DM$=0.3\,\pcc$ is observed. 
However, the linear polarization (L) is depolarized here. While our frequency resolution should enable us to measure RM up to $10^4~${\rmm}, fast RM variation of $\sim100$\,{\rmm} is most commonly seen, despite a much larger value expected from the DM variation. Linear depolarization often appears at a significant fraction of the orbit, especially during the DM variation. This is best explained by the depolarization because pulses with large varying RMs are summed together (see Methods)
The commonly observed random RM variation and linear depolarization show a highly variable magnetized environment.

\begin{figure*}
  \begin{minipage}{0.66\linewidth}
  \includegraphics[width=\linewidth]{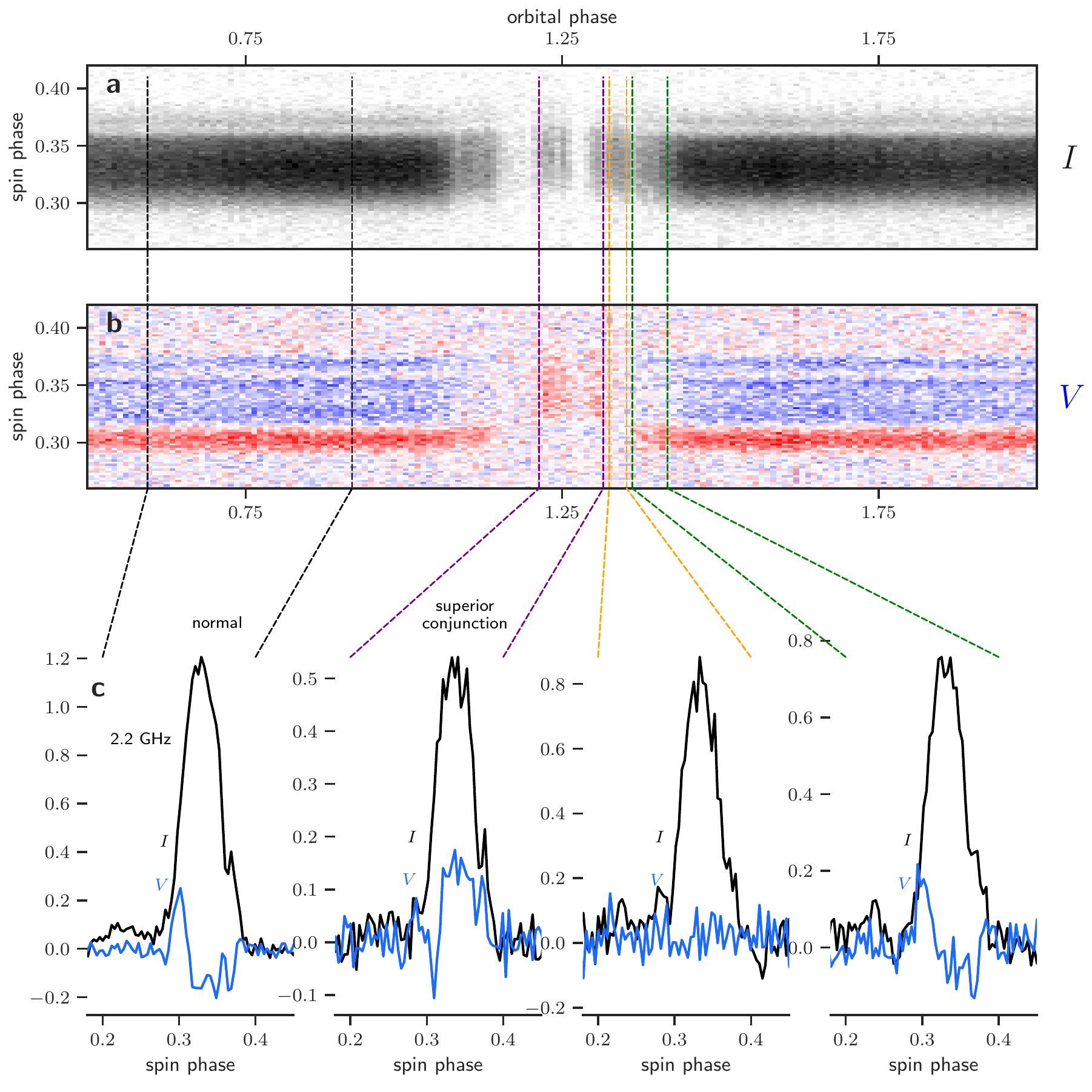}
  \end{minipage}
   \begin{minipage}{0.33\linewidth}
     \centering
  \includegraphics[width=\linewidth]{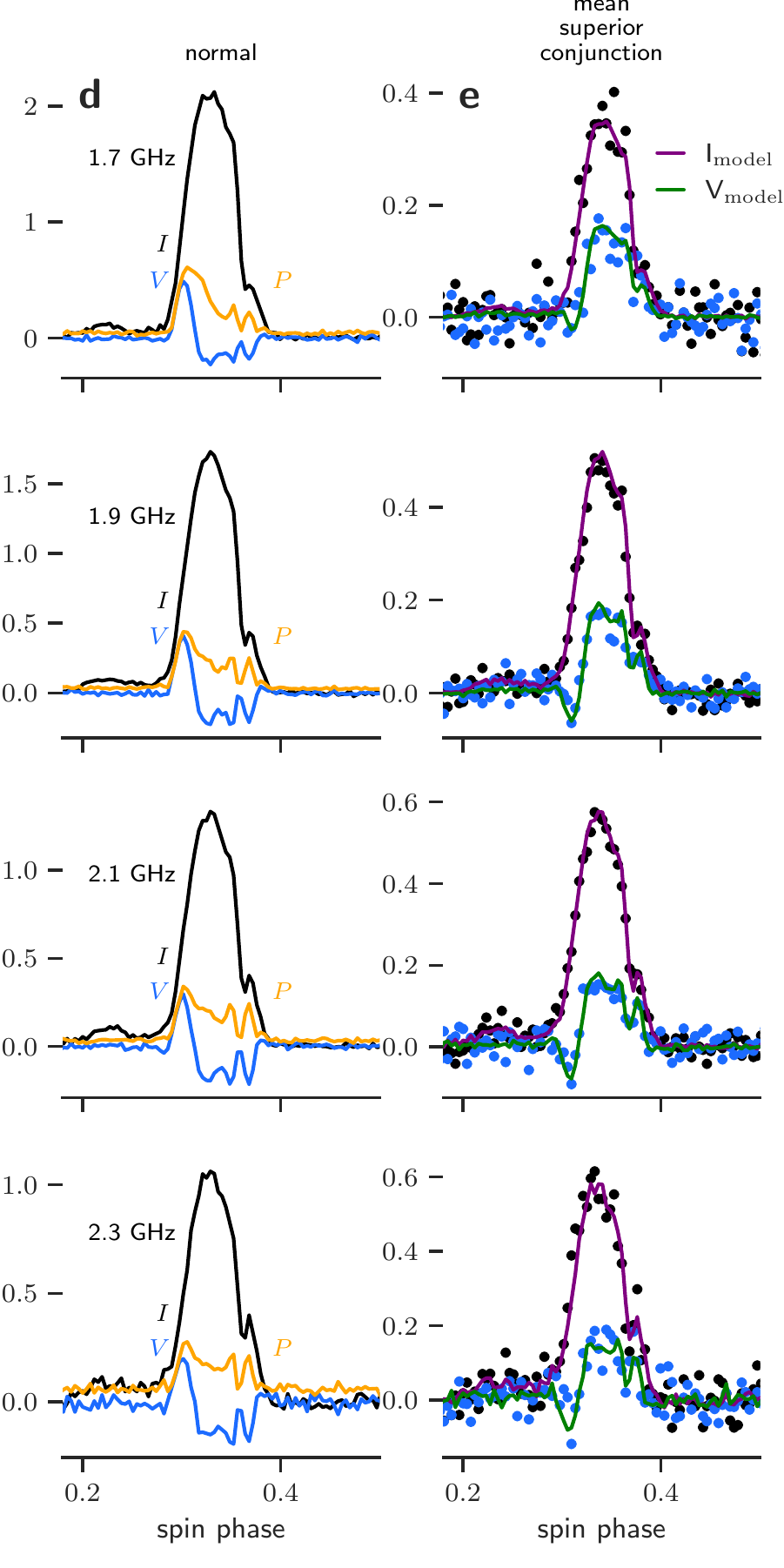}
\end{minipage}
      \caption{\textbf{Ter5A Faraday conversion near superior conjunction.} (\textbf{a,b}): Zoom-ins of the $I$, $V$ profiles in the bottom panel of Fig~1. Emission observed in the selected orbital phase range is averaged to produce the pulse profiles in panel \textbf{c}. The $V$ (blue curve) first disappears and then changes  sign when approaching  superior conjunction. (\textbf{d}): The unaffected (i.e. normal) $I$ (black), $V$ (blue) and polarization fraction $P$ (orange) against spin phase and frequency. (\textbf{e}): Observations (points) and model (lines) of the $I$ and $V$ profiles near superior conjunction averaged over three orbits. The observed profiles are reproduced by the normal profile in panel \textbf{d} going through Faraday conversion and circularly polarized absorption (see Methods).}
  \label{fig:observed_conversion}
\end{figure*}

During another observation, significant variation of circularly polarized emission $V$ was detected near superior conjunction in three consecutive orbits recorded at 2\,GHz (Figure~\ref{fig:observed_conversion}).
Near superior conjunction, $V$ has an opposite sign compared to the normal $V$ profile, slowly turning to zero and then flipping back to the normal profile as the pulsar moves away from behind the companion.
The synthetic sign reversal of $V$ across the pulse profile is a definite evidence of Faraday conversion when the radio wave adiabatically tracks a reversal in the parallel magnetic field $B_z$, an effect also known as mode coupling/tracking  \cite{Thompson1994,Melrose1995,Gruzinov2019}. 

Faraday conversion is considered an important mechanism to explain the recently observed circular polarization behavior of FRB 20201124A \cite{Xu2021,2022Kumar}. Ter5A, with which the Faraday conversion is first seen on a pulsar, has the unique advantage that the propagation effects can be isolated from intrinsic polarization behavior, as the latter is known owing to the observation away from superior conjunction. 
Propagation will affect all parts of the spin phase equally, and the complicated $V$ profile give multiple independent measurements of the parameter of the radio transfer function when compared to the unaffected $V$ profile. 
We show that the $I$, $V$ profile across frequency and spin phase can be well reproduced with the normal profile going through a reversal of circular polarization and circularly polarized absorption (Figure~\ref{fig:observed_conversion} panel d). 

\begin{figure}
  \centering
  \begin{minipage}{\linewidth}
  \includegraphics[width=\linewidth]{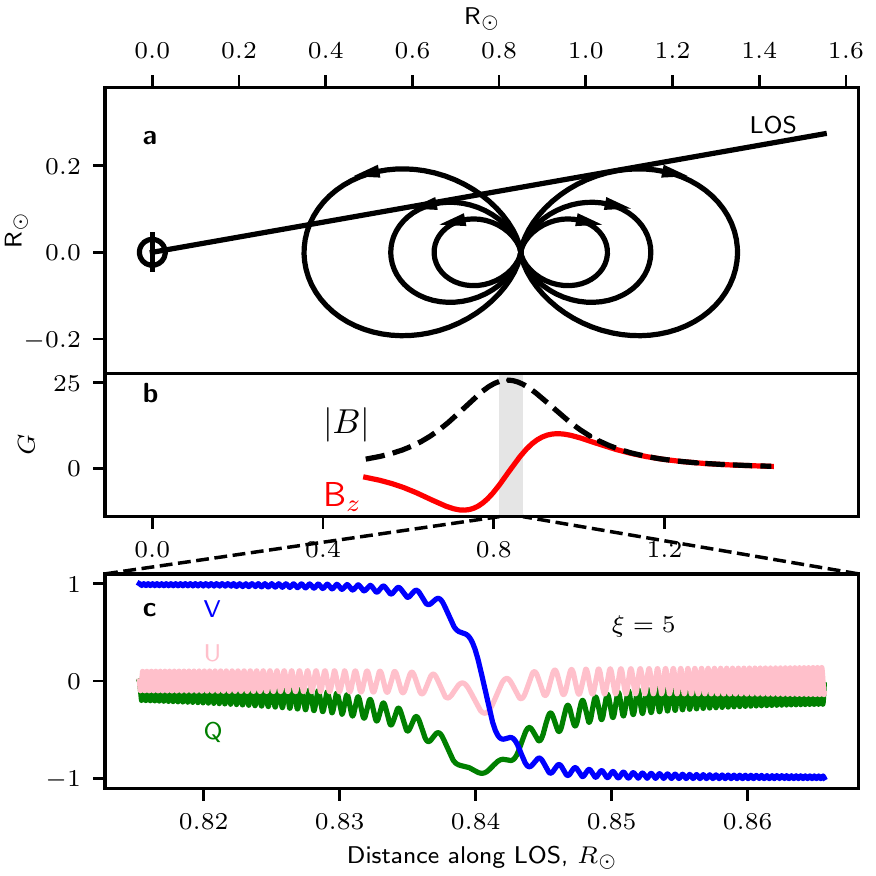}
  \end{minipage}
      \caption{\textbf{The model of Faraday conversion in the companion poloidal magnetic field.} 
      (\textbf{a}): The geometry of the system. The pulsar is located at (0,0). (\textbf{b}): The magnetic strength $|B|$ and parallel magnetic field $B_z$ in the companion magnetosphere along the line of sight (LOS). $B_z$ has changed its sign in the shaded region. (\textbf{c}) The change of Stokes Q, U, V along the line of sight (LOS) for the shaded region in (b). The $V$ changes its sign after passing the region, because the radio wave adiabatically traces the field reversal.
      For this illustrative example, we assume the companion has a dipole field with a dipole moment of $2\times 10^{32}$ $\mr{G}\, \mr{cm}^3$ (300~G surface field at the equator), orbital inclination angle $i=75^\circ$, and an electron density $n_e\approx\Delta\mr{DM}/a\approx5\times 10^{6}\mr{cm}^{-3}$, where $a$ is the orbital separation. (b,c) are shown for orbital phase 0.3 at 2~GHz. A reversal of $B_z$ will be experienced in the poloidal field as long as the LOS is not parallel to the magnetic axis. The predicted behavior of $V$ against frequency and orbital phase is shown in Extended Data Fig~\ref{fig:Voverf}.
      } 
  \label{fig:demo}
\end{figure}

The Faraday conversion is best explained with the radio wave passing through the poloidal field of the companion magnetosphere with $B>10\,G\,(\Delta \mr{DM}/0.1\,\mr{pc\,cm}^{-3})^{-1/3} (f/2\,\mr{GHz})^{-4/3}$ (see Methods). 
A reversal of parallel magnetic field $B_z$ will be stably experienced due to the geometry of the large scale poloidal field (Figure~\ref{fig:demo}). 
A plausible scenario for the circularly polarized absorption is the synchrotron-cyclotron absorption, which suggests $B\sim100$\,G given the observed increase of $V$ fraction, consistent with the lower limit from the Faraday conversion (see Methods).  
With the model, we give a qualitative prediction of $V$ against orbital phase for higher frequency (Methods, Extended data Figure~\ref{fig:Voverf}). Birefringence of lensed single pulses\cite{Bilous2019,Li2019} near superior conjunction will also be expected (see Methods). 
Both the Faraday conversion and the circularly polarized absorption indicates a highly magnetized companion. 
The change of the polarization profile against orbital phase could provide a measure of the orbital inclination angle, as the orbital geometry strongly affects the observed values of B throughout the orbit (\cite{Lyutikov2005} and see Methods). 

Magnetized companions have been observed to introduce RM variations or depolarization of linear fraction near periastron or superior conjunction \cite{1996Johnson,Crowter2020,You2018}.
Our observations of Ter5A further show that in some systems, the magnetized windy companion can introduce order-one RM variation even when the companion is almost behind the pulsar.
When the binary system is observed edge-on, RM/DM variations close to phase 0.25 will be seen; while 
if this binary system is observed near face-on, irregular RM variation of both signs would be seen depending on the intra-binary weather, which contains little information about the orbital phases. 

The irregular large RM changes are similar to what has been observed in repeating fast radio bursts (FRBs). Four out of six FRB repeaters with more than two published RMs have shown RM variations \citep{Michilli2018,Xu2021,Luo2020,2022Anna-Thomas,22Dai}. Most prominently, FRB 20190520B and FRB 20201124A have shown prominent fast irregular RM changes, which can be modelled with the existence of a companion \cite{2022Wang}. Moreover, FRB 20201124A has shown changes of circular polarization on a small fraction of the bursts, which is attributed to the polarized absorption and Faraday conversion\cite{Xu2021,2022Kumar}. We show that these effects can also appear near a companion. Moreover, the absorption mechanism responsible for the increase of $V$ fraction at Ter5A may also be responsible for the rare large $V$ observed in some FRBs \cite{2020Day,Xu2021,2022Kumar2}.
Near superior conjunction, the estimated $\gtrsim 10$\,G field from Faraday conversion and $\Delta \mr{DM}\approx 0.2 $\,\pcm yields an RM of $\gtrsim10^6$\,m$^{-2}$. This is comparable to the RM of FRB\,20121102A \cite{Michilli2018}, which has one of the largest RMs 
of any astronomical source. The limited detection of large RM variation in pulsar binaries is likely due to depolarization arising from averaging pulses with different RMs and potential scattering (see Methods). 
We will be able to observe much larger RM and RM variation if single pulses, similar to FRBs, can be detected in the binary system. 

Apart from the similarity in the observed polarimetry, there is other independent evidence suggesting the possibility that some FRBs reside in a binary system. FRB~20180916A is observed to have a 16-day period \cite{2020Chime}; while FRB~20121102A potentially has a periodicity of 160 days\cite{2020Rajwade,2021Cruces}. Binary orbit has been proposed as an origin for the long-term periodicity\cite{2020Lyutikov,2020Ioka,2021Li,2021Deng,2021Sridhar}. Additionally, the closest extra-galactic FRB~20200120E is localized to a globular cluster\cite{2021Bhardwaj,2022Kirsten} only 3.6 kpc away. 
Globular clusters host old stellar populations where traditional supernovae-formed magnetars will be too old to stay active. Even considering the dynamic formation channels of the magnetars, the merging rate require magnetic activity lifetimes longer than empirically-constrained lifetimes of Galactic magnetars \cite{2021Kremer,2022Lu}. On the other hand, binary pulsars are overabundant in globular clusters. Together, they call for consideration of the existence of a companion for a fraction of FRBs. 


\clearpage

\begin{methods}
\subsection{Data recording and reduction.}
The observations of Ter5A were recorded with the GUPPI backend on Oct 14th 2014 and the VEGAS backend on Jun 21st 2020 in a 800-MHz band centered at 1500\,MHz (L-band, session 141014) or 2000\,MHz (S-band, session 200621) in coherent dedispersion mode with time resolution of 10.24\,$\mu$s.  
The signal was dedispersed in each of 512 1.56-MHz channels with the average DM of the cluster, $\mathrm{DM}=238.0$\,pc cm$^{-3}$. The raw data were folded modulo topocentric pulse spin period and integrated within every 20\,s with \texttt{dspsr} software. The folded archives contained four Stokes parameters, 512 frequency channels, and 512 spin phase bins (corresponding to $t_\mathrm{res}=22.6$\,$\mu$s). 

The initial ephemerides for folding were obtained using observations spanning 2010-2014 
\cite{Bilous2019}. For the 2020 session, ephemerides were iteratively refined using pulse times of arrivals (TOAs) away from eclipses for seven sessions in 2019-2020. TOAs were determined using single-frequency profile templates constructed from the average uneclipsed profile in the sessions with brightest S/N. Pulsar spin period, period derivative and DM were updates using all seven observations, fitting for arbitrary phase offset between L- and S- TOAs. Subsequently, the epoch of the ascending node, DM and projected semimajor axis were updated on per-session basis using simple BT binary orbit. 

We performed full-Stokes calibration of the folded archives using standard techniques. Prior to each observation, a pulsed calibration signal was recorded which was used together with quasar B1442+101 as unpolarized flux calibrator to correct for the instrumental response of the receiver system. Polarization calibration for L-band was conducted using predetermined Mueller matrix solutions, which described the cross-coupling between orthogonal polarizations in the receivers (van Straten 2004). The Mueller matrix was determined using the PSRCHIVE task \texttt{pcm} based on observations of PSR B0450+55. For polarization calibration in S-band we assumed that the feed is ideal and consists of two orthogonally polarized receptors. Such technique was previously used for S-band observations of another MSP \cite{Bilous2015} and it proved to be adequate when comparing to more rigorous polarization calibration with predetermined Mueller matrix solutions. 

DM was measured with tempo2 using single-frequency template on a variety of timescales ranging from 10\,s to 4\,min if a subintergation had sufficient S/N of pulsed emission in each of 4 subbands. RM
was measured with rmfit on 20-s subintegrations with 256 subbands.

\subsection{RM variation and linear depolarization}
In the quiescent phase, Ter5A's radio emission has a considerable level of fractional linear polarization (see Figure~\ref{fig:observed_conversion} ``normal'' pulse profile), which allows measuring RM via Faraday rotation.
In an observation carried on 2014 with 800 MHz bandwidth centered at 1.5 GHz,
irregular fast RM changes are observed (Figure~\ref{fig:obs_overview}). Various single-integration (20~s) RM outliers are seen in different orbital phases. And at orbital phase 1.9, i.e. $\sim 120^\circ$ away from the inferior conjunction, the RM suddenly increases by $\sim 100$ {\rmm} in less than 20~s and lasts for $\sim6$\% of the orbit.
No significant dispersion measure (DM) variation $>0.01$\,\pcm\ is detected during the RM variation,
giving a lower limit of magnetic field of $\langle B\rangle=
12~\mr{mG}\, (\Delta\mr{RM}/100~\mr{rad}\,\mr{m}^{-2})/(\Delta\mr{DM}/0.01~\pcc)>12~\mr{mG}$ at orbital phase 1.9 where the pulsar is not behind the companion. The large change in RM happening in less than $\Delta t\sim20$\,s
suggests either turbulent activities at extremely short timescale, or, more likely, a small-scale structure in electron column density or magnetic field, with spatial scale $\Delta L \lesssim v\Delta t=0.01\sr$, where $v\approx800$\,km\,s$^{-1}$ is the orbital velocity assuming the neutron star mass of $1.4M_\odot$.

 Multiple regions of excess DM are observed at various orbital phases, around which the linear polarization disappears.
  This is expected from the underlying fast RM variation. To obtain enough S/N, each data point is the sum of 20~s of pulses. If the pulses have passed through an RM gradient of $\Delta \mr{RM}$ in the 20s, the linear polarization will drop to $L_\mr{obs}/L_\mr{emi}=\sin(\Delta RM \lambda^2)/(\Delta RM \lambda^2)$ of the original level. At 1.5~GHz, an RM gradient of $\sim100$~{\rmm} in the integrated pulses will reduce the linear polarization to 20\% of the original level. Therefore, RM changes greater than 100~{\rmm} in one sub-integration time will leads to depolarization of the observed linear polarization. This has nicely explained why RM variation of $\sim 100-200$~{\rmm} is most commonly seen in this system, while larger value is expected from the DM variation and the frequency resolution should enable RM measurement all the way to $10^4$~{\rmm}. The linear polarization will drop faster if there is random RM variation in the integrated pulses. With a standard deviation of $\sigma_\mr{RM}$, the linear fraction will drop as $L_\mr{obs}/L_\mr{emi}=\exp{(-2\sigma_\mr{RM}^2\lambda^4)}$\cite{1966Burn}. Therefore, the linear polarization will disappear if $\sigma_\mr{RM}\gtrsim 25~${\rmm}. Near orbital phase 0.9, the $\Delta$ DM rapidly increases to $\sim0.3\,\pcc$, suggesting a quick change of $\Delta\mr{RM}>3000\,\mr{rad/cm}^2$ given the lower limit of $\langle B\rangle$ at phase 1.9, which will certainly lead to the observed depolarization of linear polarization.
  In this case, if single FRB-like pulses can be detected from this system, orders larger RM variations, at least 3000~\rmm, will likely be seen.

  Apart from summation of pulses of different RMs, summation of scattered light passing through paths of different RMs will also lead to depolarization. This has been used to explain the decrease of polarization fraction in lower frequency for several FRBs\cite{2022Feng,2022Yang}. Estimated from the scintillation time of Ter5A\cite{Bilous2019}, the separation of scattered light at the distance of the companion due to the interstellar medium scattering should be of the order of 10~m, which will unlikely cause depolarization. However, strong plasma lensed pulses have been observed in this system\cite{Bilous2019}, suggesting small scale DM variation in the companion wind, which can also cause scattering. The separation of the scattered light $\Delta x_\mr{sct}$ can be estimated from the scattering time $\tau$ with $\Delta x_\mr{sct}=\sqrt{2ac\tau}=0.01\sr\sqrt{\tau/0.1~\mr{ms}}$. Therefore, if some pulses have scattering time longer than 0.1~ms, the separation of the scattered light will be similar to the distance the pulsar have moved in the 20~s integration time, and hence some single pulses may also experience depolarization, similar to FRBs.

\subsection{Faraday conversion in cold plasma}

In the S-band observation, when the pulsar is behind the companion, the DM increases by 0.2\,\pcm and the circular polarization appears to have a sign reversal at 2~GHz. The changes in the circular polarization indicates that the radio flux has gone through a Faraday conversion caused by the plasma and magnetic field of the companion.

In the cold plasma, assume wave propagates along $z$ direction, the polarization vector $\vp=(Q,U,V)$ change follows:
\begin{equation}
\frac{d\vp}{dz}=\vom\times\vp
\label{eq:dpdz}
\end{equation}
where $\vom=(\rhq,\rhu,\rhv)$. $\rhv$ is the Faraday rotation rate:
\begin{equation}
     \rhv=-\frac{2\pi}{c}\frac{f_p^2f_B}{f^2}\hat{B}_z
     \label{eq:rhvc}
\end{equation}
and $\rhq,\rhu$ is the Faraday conversion rate:
\begin{align}
     \rho_L=\rhq+i\rhu=-\frac{\pi}{c}\frac{f_p^2f_B^2}{f^3} (\hat{B}_x+i\hat{B}_y)^2
\label{eq:rhoL}
\end{align}
where $\hat{B}\equiv\vec{B}/B$;
$f_p=\sqrt{n_e e^2/\pi m_e}=9.0\,\mr{kHz}(n_e/\mr{cm}^{-3})^{1/2}$ is the plasma frequency, $f_B$ is the cyclotron frequency $f_B=eB/2\pi m_e c=2.8\,\mr{MHz}(B/\mr{G})$.
This procedure can be visualized as a rotation of polarization vector around the natural axis in the direction of $\vom$ in the Poincare sphere. 

Define
\begin{equation}
    	\tz \equiv 2 \frac{f}{f_B} \bz = 2 \frac{f}{f_B} \sin\alpha
    	\label{eq:tz}
\end{equation}
where $\alpha$ is the pitch angle between the wave vector and the magnetic field.

When $\tz \gg1$, i.e. $\rho_V\gg \rho_L$, the natural axis $\vec{n}=\vom/|\vom|$ is pointing towards the $V$ direction, resulting in only $Q$, $U$ rotation -- an effect known as Faraday rotation. Faraday conversion happens when the conversion rate is similar or higher than Faraday rotation rate, i.e. $\tz \lesssim 1$. In this case, the natural axis $\Omega$ is pointing close to the linear axis, and the wave modes become quasi-linear, resulting in circular/linear conversion.

As can be seen from Equation~\ref{eq:rhvc},\ref{eq:rhoL}, the Faraday rotation rate is proportional to $f^{-2}$, while the Faraday conversion rate is proportional to $f^{-3}$ for cold plasma.

In order for the Faraday conversion to happen, i.e. $\tz\lesssim 1$, $f_B\gtrsim 2f\sin\alpha$ (Equation~\ref{eq:tz}),
\begin{equation}
    B\gtrsim 1400 G \left(\frac{f}{2 \mr{GHz}}\right) \cos\alpha
    \label{eq:Bjudge}
\end{equation}
For a random pitch angle of $\alpha$, an extremely large magnetic field is required for the Faraday conversion to happen.
The phase difference between two linear eigenmodes $\theta_f$ (Faraday conversion angle) is:
\begin{align}
   \theta_f&=\int \rho_L dz
   \approx  \frac{\pi}{c}\langle\frac{f_p^2f_B^2}{2 f^3}\rangle L \\
   &\sim 10^6 \langle\left(\frac{\Delta\mr{DM}}{0.1\mr{pc}\, \mr{cm}^{-3}}\right) \left(\frac{B}{1000\,G}\right) \rangle \left(\frac{f}{2\,\mr{GHz}}\right)^{-3}
\end{align}
Given the observed DM excess of $\sim$ 0.1\,\pcm, the conversion angle is $\gg1$~rad. Therefore, if $f_B\sim f$, the circular polarization will oscillate fast against frequency, resulting in depolarized circular polarization with our frequency resolution. Moreover, it will be fine-tuning to require it to persist stable $180\degree$ conversion angle across $40\degree$ of the orbital phase. Therefore, this scenario is inconsistent with the observation (Figure~\ref{fig:observed_conversion}).

To avoid the fast change of conversion angle, we require $f\gg f_B$. In this case, the Faraday conversion will happen when the magnetic field is almost perpendicular to the LOS, i.e. $\bz=\cos\alpha \ll1$ in Equation~\ref{eq:Bjudge}. This would happen when there is magnetic reversal.

Consider a field reversal happening at $z=0$, we can Taylor expand $\rhv=\rho_V^\prime z$. Choose the $x$-axis to be the direction of perpendicular magnetic field at the reversal, the angular frequency of the rotation can be simplified as $\Omega=(\rhl, 0, \rho_V^\prime z)$.
Define
\begin{align}
	\xi &\equiv\frac{\rhl^2}{ \rho_V^\prime}\approx\frac{\pi L}{2c} \frac{f_p^2 f_B^3}{f^4}
	\label{eq:xi}
\end{align}
where $L$ is the spatial scale where the magnetic field changes direction (i.e. $|\Delta B_z|\sim |B|$). $\hat{B}_x$ is omitted in the expression because $\hat{B}_x\sim 1$ near field reversal.

Near field reversal $\rhl^\prime/\rhl \ll \rhv^\prime/\rhv$, so Equation~\ref{eq:dpdz} can be simplified as
\begin{equation}
	\frac{d\vp}{\xi d\tz}=(1,0,\tz)\times\vp
\end{equation}

Assume $\rhq$, $\rhv^\prime$ can be considered as constants near the field reversal for $-\tz_0<\tz<\tz_0$, where $\tz_0\gg 1$. Then the conversion angle  $\theta_f(\xi)$ depends only on $\xi$ \cite{Gruzinov2019}.
\begin{equation}
	 \theta_f(\xi)=\arccos (2\exp(-\pi\xi/2)-1)
\end{equation}
With $\xi\ll 1$, $\theta_f\sim \sqrt{2\pi\xi}$, while when $\xi\gg 1$, the circular polarization would flip sign.

As shown in Figure~\ref{fig:observed_conversion}, the sign of circular polarization of the pulsar flux has reversed near the orbital phase 0.25, suggesting $\xi\gg 1$ near the superior conjunction where pulsar is behind the companion. The sign reversal can persist for more than 10\% of the orbital phase, corresponding to a spatial scale of $L\sim 0.6 \sr$ at the distance of the companion. This spatial scale is of the same order as the size of the companion, therefore, we attribute the sign reversal to the large scale structure of the companion.  For a companion with a poloidal field, e.g. a dipole field, any LOS passing through the magnetosphere will experience a field reversal (Figure~\ref{fig:demo}), usually near the magnetic pole.
We can estimate the required field strength from Equation~\ref{eq:xi}.
The dispersion measure (DM) changes by $\sim0.1\pcc$ during the conversion, giving a mean density of $n_e\approx N_e/a=5\times 10^6$\,cm$^{-3}$; where $a=0.85 R_\odot$ is the orbital separation. It corresponds to plasma frequency $f_p=0.02$\,GHz. We use the spatial scale $L$, where the conversion persists, as an approximate for the scale where magnetic field changes direction. Therefore we require the average $B\gtrsim 10$\,G in the companion's magnetosphere.

The nature of the 0.089 $M_\odot$ companion is unknown, however, whether it is an M dwarf or white dwarf, it is common to have a large surface magnetic field\cite{19Shulyak_mdb}.
Moreover, a magnetic field of the order of 10\,G at the contact discontinuity is expected from the pressure equilibrium with the pulsar wind. The companion has been seen to eclipse the pulsar for a whole orbit, requiring a large pressure in the companion wind to withhold the pulsar wind. The pressure in the pulsar wind at the distance of the companion can be estimated from the spin down energy $P_\mr{pw}=\pi I \dot{P}/P^3 a^2 c$, where I$\sim10^{45}$\,g\,cm$^2$ is the pulsar's moment of inertia, $\dot{P}$ is the period derivative, and $c$ is the speed of light. The observed period derivative of Ter5A is $\dot{P}= -1.5 \times 10^{-20}$\,s/s which is dominated by the acceleration in the globular cluster. Given the location of the pulsar in the Terzan 5 cluster, the maximum contribution of the $\dot{P}$ from the cluster dynamic is $\approx2\times 10^{-19}$~s\,s$^{-1}$, and both the mean and median value estimated from all the reasonable cluster models are greater than half of this value \cite{Phinney1992}. Taking the cluster contribution to be $\dot{P}\sim 10^{-19}$\,s\,s$^{-1}$, the pressure of the pulsar wind at the distance companion is $P_\mr{pw}\sim 2$\,erg\,cm$^{-3}$. To balance the pulsar wind with magnetic pressure $P_\mr{pw}=B_\mr{cp}^2/8\pi$, a field strength of $B_\mr{cp}\sim7$\,G in the contact discontinuity would be expected. Therefore, it is reasonable to expect the magnetic field in companion magnetosphere to be greater than this value, and hence satisfy the condition for the Faraday conversion scenario detailed above.

\subsection{Modeling the IV spectrum}
As shown in Figure~\ref{fig:observed_conversion},
the change of circular polarization at the top of the band ($\sim2.3$\,GHz) roughly resembles the normal $V$ profile with an opposite sign. However, in the lower half of the band ($\lesssim 2$\,GHz), in additional to the flip of the sign, the V profile is shifted systematically towards the positive value. Selecting orbital phases far from the superior conjunction, where stable RM has been measured, we can measure the normal linear polarization $L$ of the pulsar, and hence can calculate the total polarization $P=\sqrt{L^2+V^2}$ (shown in orange line in left panel d at Figure~\ref{fig:observed_conversion}). Around the spin phase 0.35, the circular polarization fraction near superior conjunction exceeds the total polarization fraction in the normal profile, and this trend becomes more obvious as the frequency decreases. At 1.7~GHz, the total polarization fraction $P/I$ at spin phase 0.35 is only 15\% in the normal profile; however, near the superior conjunction $V/I$ is around 50\% at the same spin phase, suggesting $P\ge50$\%, i.e. the total polarization fraction has changed. As shown in Equation~\ref{eq:dpdz}, the total polarization fraction $P$ is a conserved quantity in both Faraday rotation and Faraday conversion. The change of total polarization fraction against frequency, accompanied by the decrease of flux, is best explained by circularly polarized attenuation. Therefore, the radiative transfer equation (Equation~\ref{eq:dpdz}) becomes
\begin{gather}
\frac{d\vs}{ds}= - R \vs \\ 
R=
  \begin{pmatrix}
\eta &  0 & 0 & \eta_V \\
0 &  \eta & \rho_V & 0 \\
0 & -\rho_V & \eta & \rhq \\
\eta_V & 0 & -\rhq & \eta
\end{pmatrix} \nonumber
\label{eq:transfer}
\end{gather}
where $\vs=(I,Q,U,V)$; $\eta$, $\eta_V$ are the isotropic, and circular absorption.
The circular absorption happening during the adiabatic field reversal are largely cancelled. For the circular absorption happening before/after the conversion, the integrated form will be:
\begin{gather}
\begin{pmatrix}
I^\prime\\V^\prime
\end{pmatrix}=e^{-\tau}
\begin{pmatrix}
\cosh{\tau_v} & -\sinh{\tau_v} \\
-\sinh{\tau_v} & \cosh{\tau_v}
\end{pmatrix}
\begin{pmatrix}
I \\
c\,V
\end{pmatrix}
\label{eq:fit}
\end{gather}
where $\tau=\eta L$, $\tau_v=\eta_v L$ are the optical depth, $c=\cos\theta_f$ is the conversion angle.

The optical depth should be frequency dependent, there we parameterize them as $\tau=A (f/f_0)^{-\alpha}$, $\tau_v=A_v (f/f_0)^{-\alpha_v}$, where $f_0=2$\,GHz.
Near the superior structure, the mode tracking is visually complete at the top of the band, therefore we fix $c=-1$ for all of the frequencies. Using the $I$, $V$ at the normal phase as template $I$, $V$, we fit for the $I^\prime, V^\prime$ near the superior structure by varying $A, \alpha, A_v, \alpha_v$. The best fit parameters are $A=1.08\pm0.01, \alpha=-3.5\pm0.1, A_v=-0.21\pm 0.01, \alpha_v=-3.7\pm0.4$, where the errorbars are 1 $\sigma$ error. The fitted curves are shown in Figure~\ref{fig:observed_conversion} panel d right column as solid lines.

\subsection{Synchrotron-cyclotron absorption}
Given the large magnetic field required for the Faraday conversion, the synchrotron-cyclotron absorption is able to account for the circularly polarized absorption and varying absorption index at different orbital phase.

In the classic synchrotron limit, assume the relativistic electron following an isotropic powerlaw distribution $n_r(\gamma)\propto\gamma^{-p}$ for $\gamma_\mr{min}<\frac{3}{2} f_B \sin \alpha <\gamma_\mr{max}$, the isotropic and circular absorption index $\eta,\eta_v$ can be estimated with \cite{1969Sazonov}:
\begin{align}
\eta=&\frac{0.08}{\sr^{-1}} \frac{n_r}{[\mr{cm}^{-3}]}\frac{(p-1)}{\gamma_\mr{min}^{1-p}}\nonumber \\
&\Gamma(\frac{3p+2}{12})\Gamma(\frac{3p+22}{12}) \Big(\frac{3f_B\sin\alpha}{f}\Big)^{p/2+1} \frac{[\mr{GHz}]}{f} \nonumber \\
\eta_V=&\frac{0.03}{\sr^{-1}} \frac{n_r}{[\mr{cm}^{-3}]}\frac{(p-1)}{\gamma_\mr{min}^{1-p}} \\
&\frac{p+3}{p+1}\Gamma(\frac{3p+7}{12})\Gamma(\frac{3p+11}{12}) \Big(\frac{3f_B\sin\alpha}{f}\Big)^{p/2+3/2} \frac{[\mr{GHz}]}{f} \nonumber
\end{align}
Therefore $\tau_v=\eta_V L\propto f^{-p/2-5/2}$ has steeper spectrum compared to the isotropic optical depth $\tau=\eta L\propto f^{-p/2-2}$. The ratio $\eta_v/\eta \approx 0.6 \sqrt{f_B\sin\alpha/f}$.

In the methods, we have fitted $\tau=A (f/f_0)^{-\alpha}$, $\tau_v=A_v (f/f_0)^{-\alpha_v}$ for the pulsar flux. Near the superior conjunction, the fitted $\alpha,\alpha_v$ are consistent with $p=3\pm0.2$. The ratio $\tau_v/\tau=A_v/A\approx0.2$ at 2\,GHz requires $B\sin\alpha\sim100$\,G. Assume $\gamma_\mr{min}=2$ and $L=0.6\sr$, the required relativistic electron density for the eclipse is $n_r=20$\,cm$^{-3}$ near superior conjunction. The required magnetic field $B\sim100$\,G is consistent with the required $10 \mr{G}\lesssim B<700$ G from the observed Faraday conversion.

The relativistic electrons also inevitably introduce linear absorption $\eta_L$ and Faraday conversion $\rho_L^r$, however, given $n_r\ll n_e$ and $f_B<f$, these two terms will be much smaller than the Faraday rotation and conversion introduced by cold plasma (Equation~\ref{eq:rhvc},\ref{eq:rhoL}) and hence their effect will not be visible.

\subsection{Constraining the orbital inclination angle}
The stable conversion indicates that the LOS is passing through the magnetosphere of the companion, where a dipole field is expected. A wide-band observation would provide constraints on the excess DM at different orbital phase. In this way we could model the magnetic strength variation against orbital phase, and by comparing with the dipole magnetic field, we would be able to constrain the orbital inclination angle.

For a dipole field,
\begin{equation}
	|B|=|M|\frac{(1+3\cos^2\theta)^{1/2}}{r^3}
\end{equation}
Here $|M|=B_0r_0^3$ is the dipole moment, and $B_0$ is the surface field at the equator, $r_0$ is the radius of the companion. Despite the weak dependence (less than a factor of two) on polar angle $\theta$, the change of $|B|$ mainly depends on $r^{-3}$.

For a binary orbit at an inclination angle $i$ (where $i=90\degree$ is edge on), and an orbital phase $\Psi=2\pi(\Phi-0.25)$ (when $\Psi=0$ the pulsar is behind the companion), the LOS has the shortest radial distance to companion $r_\mr{min}=d \sqrt{1-\sin^2 i \cos^2 \Phi}$ (happening at a distance $R$ from the pulsar $R=d\sin i \cos \Phi$), where d is orbital separation.
When the orbit is close to edge on $i\sim 90\degree$, $r_\mr{min}\sim d \sin \phi $, the field strength changes fast against orbital phase $B\propto r^{-3}\propto \sin^{-3} \Phi$, while for a face-on orbit $i\sim0\degree$, the field strength is independent of the orbital phase. Therefore, the relative field strength against orbital phase mainly depends on the inclination angle with mild dependence on the polar angle. Extended Data Figure~\ref{fig:Boverphi} shows the relative field strength against orbital phase at different inclination angle when the magnetic pole aligned with the orbital axis.
Comparing $i=85\degree$ to $i=75\degree$, the relative strength of $B_{0.35}/B_{0.25}$ reduced by an order. Therefore, measuring the field strength against orbital phase will provide a great constraint on the inclination angle. We defer the detailed modeling to a future paper.
%

\subsection{Observation prediction}
In the scenario of Faraday conversion in the slow field reversal, for the majority of the paths, the wave is going through Faraday rotation. We can estimate the RM given the estimated magnetic field:
\begin{equation}
  \frac{<\Delta\mr{RM}>}{[\mr{rad}~m^{-2}]}\sim \frac{<\Delta\mr{DM}>}{[\mr{pc/cm}^{-3}]}\frac{B}{[\mu G]}
\end{equation}
Given the $\Delta \mr{DM}$ of around 0.2\,pc/cm$^3$, the $\Delta \mr{RM}$ would be $10^6$~{\rmm} for 10-G magnetic field. The large RM would lead to a delay $\tau_F$ between the two circular polarization, which may show up in the stokes V profile:
\begin{equation}
    <\tau_F>=\frac{4<f_{B_\parallel}>}{f}<\tau_p>
\end{equation}
where $\tau_p$ is the dispersive delay, which is $\sim0.2$\,ms. The $\tau_F$ would be $\sim10\,\mu$s for 10-G field. For B strength of $10-100$\,G, this may lead to a suppression of stokes in the phase profile where V switches sign and have order 1 modulation in the timescale of $\sim 100\,\mu$s.

Observations of single pulses near superior conjunction will be the smoking gun to verify the magnetic strength.
Plasma lensing events have been seen in Ter5A \cite{Bilous2019}, where single pulses can occasionally be magnified significantly. We will be able to measure the RM value of the single pulses given enough frequency resolution. Moreover, the presence of a large magnetic field will result in the left and right circular polarization of the pulse being magnified at different frequency\cite{Li2019}. The lensing profile will be offset by a cyclotron frequency $f_c=2.8\,\mr{MHz}(B/\mr{G})$, which is $\sim30$\,MHz for $B=10$\,G and $\sim0.3$\,GHz for $B=100$\,G, and hence should be measurable.

Last but not least, the observation suggests that the parameter quantifying Faraday conversion $\xi\gg1$ at 2\,GHz (Equation~\ref{eq:xi}) and hence $|B|\gtrsim 10$\,G near the superior conjunction. It also suggests $f_B\ll f$ for most of the path, and hence $|B|\ll700$\,G.
Since $\xi\propto f^{-4}$ has strong dependence on frequency, the transition from full conversion to negligible conversion will happen quickly.  Given $B\ll700$\,G, and the conversion would stops happening with $\xi\ll1$, the transition from full conversion to negligible conversion must happen between 2\,GHz to 20\,GHz (referred to as transition frequency $f_T$).
$\xi\propto f_B^3\propto B^3$ is very sensitive to the magnetic field strength. At different orbital phase, the LOS is passing through different part of the magnetosphere, therefore, we expect the transition frequency $f_T$ to decrease as the pulsar moves away from the superior conjunction. We have observed $f_T=2$~GHz at around orbital phase 0.35, where $V$ starts to change sign. At higher frequency we will be able to observe $f_T$ closer to $\Phi=0.25$. Therefore, at higher frequency, there will be a narrower window in the orbital phase where $V$ changes sign.

To demonstrate the model and the prediction, we configured a toy model with a magnetized companion with a dipole moment of $2\times 10^{32}$ $\mr{G}\, \mr{cm}^3$ (300~G surface field at the equator, 0.12~$\sr$ companion radius). With orbital inclination angle $i=75^\circ$, and an electron density estimated with $n_e\approx\Delta\mr{DM}/a\approx5\times 10^{6}\mr{cm}^{-3}$, where $a$ is the orbital separation, we can reproduce the Faraday conversion behavior of the pulsar against orbital phase for 2.3~GHz and show the behavior of $V$ at higher frequency (Extended Data Fig~\ref{fig:Voverf}).

\end{methods}
\clearpage

\subsubsection*{Data availability}
Data are available at request.
\subsubsection*{Code availability}
\noindent
\textsc{DSPSR} (\url{http://dspsr.sourceforge.net})

\noindent
\textsc{PSRCHIVE} (\url{http://psrchive.sourceforge.net})

\begin{addendum}
\item
AVB is supported by the European Research Council under the European Union's Seventh Framework Programme (FP/2007-2013)/ERC Grant Agreement No. 617199 (`ALERT') and by Vici research programme `ARGO' with project number 639.043.815, financed by the Dutch Research Council (NWO).
YPY is supported by NSFC grant No. 12003028 and the China Manned Spaced Project (CMS-CSST-2021-B11).
SMR is a CIFAR Fellow and is supported by the NSF Physics Frontiers Center awards 1430284 and 2020265.
The National Radio Astronomy Observatory is a facility of the National Science Foundation operated under cooperative agreement by Associated Universities, Inc. The Green Bank Observatory is a facility of the National Science Foundation operated under cooperative agreement by Associated Universities, Inc.
\item[Author Contributions]
DZL modelled and interpreted the data, and prepared the majority of the manuscript. AVB performed data pre-processing, calibration and intermediate measurements. SMR led the data acquisition and discovered the variability of circularly polarized emission. RM and YPY helped to revise the draft and provided valuable comments.
\item[Competing Interests] The authors declare no
competing financial interests.
\item[Correspondence] Requests for materials should be addressed to D.~Z.~Li (E-mail:dongzili@caltech.edu)\\
\end{addendum}

\bibliography{main.bbl}
\clearpage

\setcounter{figure}{0}
\setcounter{table}{0}
\captionsetup[table]{name={\bf Extended Data Table}}
\captionsetup[figure]{name={\bf Extended Data Figure}}

\begin{figure}
  \centering
  \includegraphics[width=\linewidth]{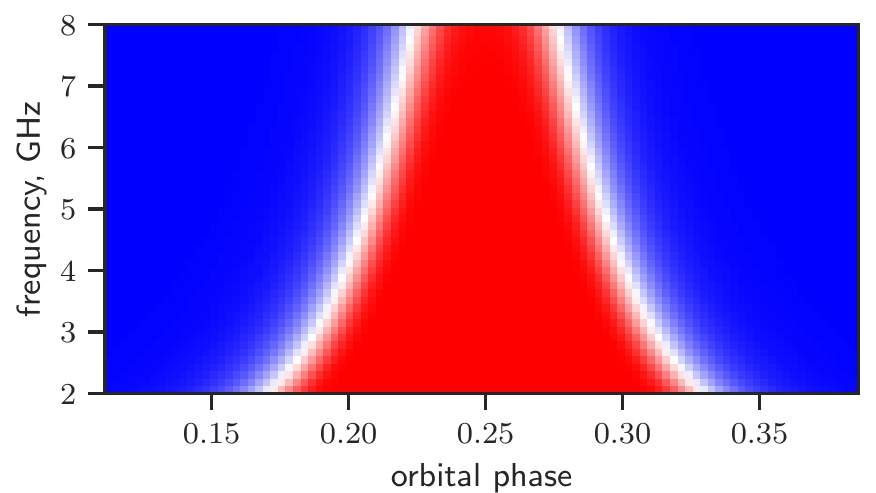}
      \caption{Predicted $V$ change against orbital phase at higher frequency with the model in Fig~\ref{fig:demo}. Here we demonstrate the $V$ behavior for a single spin phase, with red and blue representing different signs. Around the superior conjunction, $\phi=0.25$, $V$ will experience Faraday conversion resulting in the change of the sign. The window for Faraday conversion will become narrower at higher frequency. The modeling of circularly polarized absorption has more uncertainty due to the unknown distribution of mildly relativistic electron against orbital phase. However, this absorption is observed to decrease fast against frequency with $\tau_v\propto f^{-4}$, and become negligible at $f=2.3$~GHz. Therefore it is safe to ignore it at higher frequency.}
  \label{fig:Voverf}
\end{figure}

\begin{figure}
  \centering
  \includegraphics[width=\linewidth]{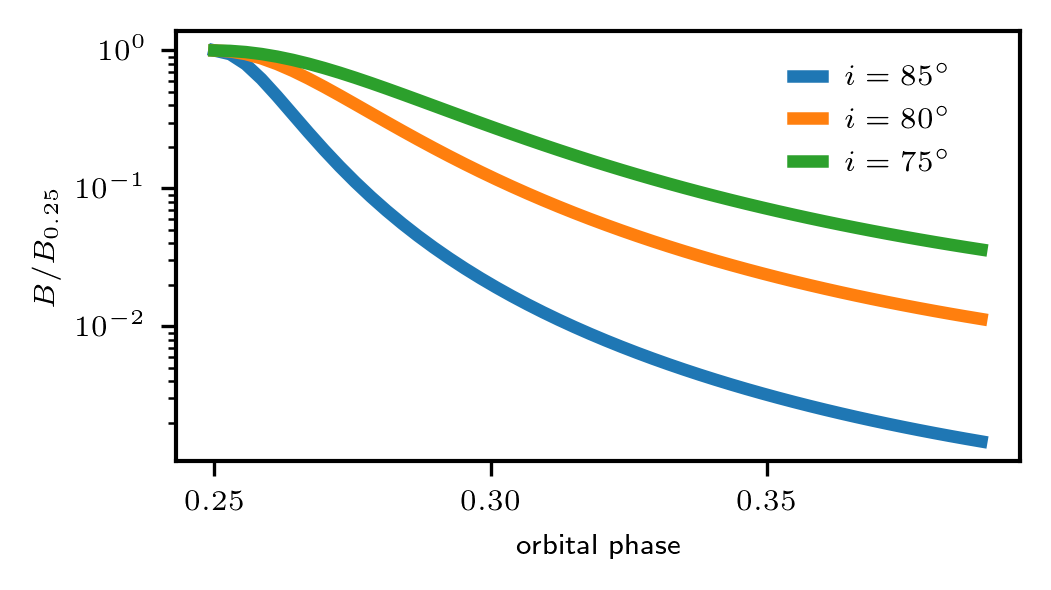}
      \caption{The peak magnetic field encountered for LOS at different orbital phases compared to the one at orbital phase 0.25. The larger the inclination angle $i$ is, the faster the field strength changes against orbital phases. Here we assume the LOS has passed through a dipole companion magnetosphere with magnetic axis aligned with the orbital axis. The result will only have minor changes with other magnetic axis alignment.}
  \label{fig:Boverphi}
\end{figure}

\end{document}